\documentclass[12pt]{article}

\usepackage{amsmath}
\usepackage{amssymb}
\usepackage{amsthm}
\usepackage{array}
\usepackage{bbold}
\usepackage{booktabs}
\usepackage[font=small,labelfont=bf]{caption}
\usepackage{color}
\usepackage{colortbl}
\usepackage{dsfont}
\usepackage{eurosym}
\usepackage{enumitem}
\usepackage{float}
\usepackage{footmisc}
\usepackage{geometry}
\usepackage{graphicx}
\usepackage{hyperref}
\usepackage{lscape}
\usepackage{mathptmx}
\usepackage{multicol}
\usepackage{multirow}
\usepackage{natbib} 
\usepackage{setspace}
\usepackage{subfig}
\usepackage[table]{xcolor}
\usepackage[normalem]{ulem}

\definecolor{lightgray}{rgb}{0.75, 0.75, 0.75}

\begin{document}

\title{Statistical enhanced learning for modeling and prediction tennis matches at Grand Slam tournaments}

\author{N. Buhamra$^1$ \and A. Groll$^2$}

\date{%
    \small{$^1$Department of Statistics, TU Dortmund University, Vogelpothsweg 87, 44221 Dortmund, email: \texttt{nourah.buhamra@tu-dortmund.de}\\%
    $^2${\em Corresponding author}, Department of Statistics, TU Dortmund University, Vogelpothsweg 87, 44221 Dortmund, email: \texttt{groll@statistik.tu-dortmund.de}    }\\[2ex]
}	
\maketitle

\setcounter{tocdepth}{3}
\setcounter{secnumdepth}{3}

\abstract{\noindent In this manuscript, we concentrate on a specific type of covariates, which we call statistically enhanced, for modeling tennis matches for men at Grand slam tournaments. 
Our goal is to assess whether these enhanced covariates have the potential to improve statistical learning approaches, in particular, with regard to the predictive performance. For this purpose, various proposed regression and machine learning model classes are compared with and without such features. To achieve this, we considered three slightly enhanced variables, namely elo rating along with two different player age variables.
This concept has already been successfully applied in football, where additional team ability parameters, which were obtained from separate statistical models, were able to improve the predictive performance.

In addition, different interpretable machine learning (IML) tools are employed to gain insights into the factors influencing the outcomes of tennis matches predicted by complex machine learning models, such as the random forest. Specifically, partial dependence plots (PDP) and individual conditional expectation (ICE) plots are employed to provide better interpretability for the most promising ML model from this work. Furthermore, we conduct a comparison of different regression and machine learning approaches in terms of various predictive performance measures such as classification rate, predictive Bernoulli likelihood, and Brier score. This comparison is carried out on external test data using cross-validation, rolling window, and expanding window strategies.}

\vspace*{0.2cm}


\section{Introduction}

In recent years, various methodologies for statistical and machine learning based modeling of tennis matches and tournaments have emerged, and the existing methods for predicting the probability of winning matches in tennis have been expanded. Moreover, there is potential to calculate winning probabilities for an entire tournament when all individual matches can be predicted.

Recently, machine learning (ML) models have been employed to predict the outcomes of tennis matches. \citet{Somboonphokkaphan2009tennis} introduced a method utilizing match statistics and environmental data to predict winners using a Multi-Layer Perceptron (MLP) with a back-propagation learning algorithm. MLP, a type of Artificial Neural Network (ANN), is effective for solving real-world classification problems and predicting outcomes, especially when handling large databases with incomplete or noisy data. \citet{Whiteside2017monitoring} proposed an automated stroke classification system to quantify hitting load in tennis, using machine learning techniques like a cubic kernel support vector machine. \citet{Wilkens2021sports} expanded previous research by applying various ML techniques, including neural networks and random forests, with extensive datasets from professional men’s and women’s tennis singles matches. Despite these efforts, he found that the average prediction accuracy does not exceed~70\%.

\cite{sipko2015machine} predicted match winners based on the probability of winning serve points, which subsequently indicates the overall probability of winning the match. They extracted 22 features from historical data, including abstract features like player fatigue and injury, and optimized models that outperformed Knottenbelt’s Common-Opponent model using ML approaches such as artificial neural networks (ANNs). They suggest that ML-based techniques can innovate tennis betting, noting that ANNs generated a 4.35\% return on investment, a 75\% improvement in the betting market. Moreover, \cite{gao2021random} developed a model that predicts tennis match outcomes with over 80\% accuracy, surpassing predictions based on betting odds alone, and identifying serve strength as a crucial predictor. Their model used a random forest classifier, highlighting the importance of simple models even in the age of deep learning. Their comprehensive data set, compiled from ATP data from 2000 to 2016, includes a variety of features capturing physical, psychological, court-related, and match-related variables. Finally, a comprehensive overview of modeling and predicting tennis matches at Grand Slam tournaments by different regression approaches has been presented in \citet{buhamra:2024}.

The main focus of this manuscript, however, is to analyze, whether so-called enhanced covariates have the potential to improve
statistical and machine learning approaches, in particular, with regard to predictive performance.
Generally, in recent years, there has been a growing interest in feature engineering.
Effective feature engineering plays a crucial role in enhancing the performance of machine learning models by identifying and capturing relevant patterns and relationships within the data. This enables models to improve their predictive accuracy and extract meaningful insights from the data. For instance, \cite{felice2023statistically} introduce the concept of Statistically Enhanced Learning (SEL), a formalization framework for existing feature engineering and extraction tasks in ML. This approach has the potential to address challenges in ML tasks by optimizing feature selection and representation.

For example, in the context of modeling soccer, \cite{ley2019ranking} proposed a model to estimate flexible, time-varying team-specific ability parameters. The resulting estimates were then added to the set of (conventional) features in a random forest model, which turned out to be quite successful for predicting the FIFA World Cup 2018 in \cite{groll2019hybrid}.

When using modern and complex ML models, another important aspect is interpretability of the fitted model. Hence, several studies have been conducted in the field of understanding and interpreting complex (black box-type) ML models. 
For example, \cite{auret2012interpretation} used variable importance measures, directly generated by the random forest models, and partial dependence plots, indicating that random forest models can reliably identify the influence of individual variables, even in the presence of high levels of additive noise.

Moreover, \cite{goldstein2015peeking}  present both individual conditional expectation (ICE) plots and classical partial dependence plots (PDPs) on three different real data sets, namely depression clinical trial, white wine and diabetes classification in Pima Indians. They demonstrate how ICE plots can shed light on estimated models in ways PDPs cannot. Accordingly, ICE plots refine the PDP by graphing the functional relationship between the predicted response and the feature for individual observations. In particular, ICE plots highlight the variation in the fitted values across the range of a certain selected covariate, suggesting where and to what extent heterogeneities might exist.

More generally, \cite{molnar2023relating} investigated the relationship between  Partial Dependence Plots (PDPs) and Permutation Feature Importance (PFI) methods in understanding the data generating process in machine learning models. They explored how these two techniques can provide complementary insights into the importance and effects of features on model predictions. Consequently, they formalize PDP and PFI as statistical estimators representing the ground truth estimands rooted from the data generating process. Their analysis reveals that PDP and PFI estimates can deviate from this ground truth not only due to statistical biases but also due to variations in learner behavior and errors in Monte Carlo approximations. To address these uncertainties in PDP and PFI estimation, they introduce the learner-PD and learner-PFI approaches, which involve model refits, and propose corrected variance and confidence interval estimators.

Unlike traditional black-box models, interpretable machine learning (IML) models aim to provide insights into the decision-making process, enabling users to make informed decisions and understand the implications of model outputs. Therefore, in this work, we focused on IML models, such as partial dependency plots (PDP; \citealp{friedman2001greedy}) and Individual Conditional Expectation (ICE; \citealp{apley2020visualizing}), to make our employed random forest model more interpretable. Additionally, we demonstrate how feature engineering techniques can be applied in the context of sports analytics to enhance predictive modeling and gain insights from sports data. Specifically, we examine the use of covariates such as \textit{Elo}, \textit{Age.30} and \textit{Age.int}, which all are not directly available, but are obtained from either a separate modeling approach (\textit{Elo}) or via meaningful mathematical transformations (\textit{Age.30}, \textit{Age.int}), in order to enhance statistical models for  tennis. For this application, various regression and machine learning approaches were considered, including linear regression, spline models, and random forest. These approaches were then compared based on various performance measures within an expanding window strategy for analyzing tennis data from Grand Slam tournaments. To conduct this analysis, a dataset was compiled using the R package \texttt{deuce} \citep{Kovalchik2018deuce}. This data set included information on 5,013 matches from men's Grand Slam tournaments spanning the years 2011 to 2022. Several potential covariates were considered, including the players' age, ATP ranking and points, odds, elo rating, as well as two additional age variables. These additional age variables were designed to reflect the ``optimal" age range of a tennis player, which is typically between 28 and 32 years \citep{Weston:2014}.

The remainder of the article is structured as follows. Section~\ref{sec:two} briefly introduces the data set and defines the objectives of this work. Then, in Section~\ref{sec:three}, different modeling approaches are introduced, including classical regression approaches and ML methods such as random forest. Besides, some interpretable machine learning techniques like partial dependence plots (PDP) and individual conditional expectation (ICE) plots are discussed, and various performance measures are defined. In Section~\ref{sec:five}, the modeling approaches are compared in terms of various performance measures, using an expanding window strategy. Additionally, interpretations for different model classes considering enhanced covariates and IML techniques are provided. Finally, Section~\ref{sec:six} summarizes the main results and gives a final overview.

\section{Data}\label{sec:two}

Next, we shortly introduce our data set, which was compiled using the R package \texttt{deuce} \citep{Kovalchik2018deuce}, and contains information on 5,013 matches from men's Grand Slam tournaments from 2011 to 2022. It was already used in \citet{buhamra:2024} and \citet{buhamra:2025}, and contains the following variables.

\begin{description}
	\item[\it Victory:] A dummy variable indicating whether the first-named player won the match (1: yes, 0: no), used as the response variable in all models.
 \end{description}
 
 \subsubsection*{Conventional covariates}

 The following three covariates are conventional covariates, which could be extracted more or less directly from public sources. They are all incorporated in our final data set in the form of differences, i.e.\ for each feature the value of the 2nd player is subtracted from the one of the 1st player.

 \begin{description}
	\item[\it Age:] A metric predictor collecting the age of the players in years. Note that players' ages were not given directly and had to be deduced from the player's date of birth as well as the date of the respective match. 

	\item[\it Rank:] For each player, the rank at the start of the respective tournament was collected. The position in the ranking is based on the ATP ranking points.

	\item[\it Points:] The ATP world ranking points are awarded for each match won per tournament. Wins in later rounds of a tournament are valued higher than wins in the first rounds of a tournament. Points earned in a tournament expire after 52 weeks.
 \end{description}

\subsubsection*{Enhanced covariates}

Next, we introduce three covariates which we call enhanced, as they are not directly available, but are obtained from either a separate modeling approach ({\emph{Elo}) or via meaningful
mathematical transformations (\emph{Age.30, Age.int}), in order to enhance statistical models
for tennis. Again, also these features are all incorporated as differences (value of 2nd player minus value of 1st player).

\begin{description}
	\item[\it Elo:] For each player, the overall Elo rating before a certain match is collected. The idea of the Elo rating system is that one can more accurately track and predict player performances over time, taking into account the relative strength of opponents and the outcomes of matches. Each player starts with an initial rating (often set around 1500 points, though this can vary depending on the specific implementation). After each match, the players' ratings are updated based on the match outcome. If a higher-rated player wins, they gain fewer points than if a lower-rated player wins. The amount of points exchanged in a match depends on the difference in ratings between the two players. 
 Hence, this covariate is considered ``enhanced'' as it involves complex calculations and provides a more delicate measure of player performance. Further details on the Elo rating in tennis can be found in \cite{angelini2022weighted} and \cite{vaughan2021well}

	\item[\it Age.30:] This variable is given by the absolute distance between the age of the players and reference age 30. This is based on the assumption that the standard \emph{Age} variable introduced above does not contain enough information. For example, while a 25-year-old player typically has an age-advantage over a 20-year-old one, a 40-year-old player rather has a disadvantage over a 35-year-old one; and, in both cases, the age difference between the two players equals 5 years. Following \citet{Weston:2014}, who argued that the optimal age of tennis players is between 28 and 32 years, we chose the mid-point as the reference age.
	
	\item[\it Age.int:] This feature is based on the assumption that the optimal age of a tennis player lies within the interval $[28;32]$. Then, the smaller distance to the limits of this interval was derived, i.e.\ for players younger than 28 the distance to 28 was calculated and for players older than 32 the distance to 32 was calculated. For players between 28 and 32 the distance was set to 0. 	
\end{description}
A detailed description of the variables included in the data set can be found in Section~2 of \citet{buhamra:2024}  and  \cite{buhamra:2025} .

Note at this point that the data set does not include matches in which one of the two players gave up or was unable to compete, e.g.\ due to injury, such that the other player won without actually playing the match. These matches do not contain any relevant information for the present analysis and, hence, in order not to distort the results, are excluded. Moreover, the data set does not contain any missing values. 

Based on this data set, the best possible statistical enhance learning model for predicting tennis matches at Grand Slam tournaments is sought. Also, it will be examined whether a machine learning approach, namely a random forest, incorporating enhanced statistical covariates, is more powerful in predicting tennis matches compared to classical regression approaches that also incorporate the corresponding enhanced covariates. Then, for all proposed modeling approaches, including machine learning and classical regression methods, optimal models are determined based on certain performance measures in terms of an expanding window prediction approach. 
Here, our focus will be only on the expanding window strategy, which reveals a clear winner model, but also other technique such as leave-one-tournament-out cross-validation and a rolling window approach have been considered. The results for those approaches can be found in the appendix. Our main objectives are (i) to quantify how the predictive performance can be improved by incorporating enhanced variables, and (ii) to provide better interpretations for the random forest model using IML tools such as PDP plots and ICE plots.

\section{Methods} \label{sec:three}
In the following, first the statistical and machine learning methods used in this work are briefly introduced in Section~\ref{sec:regression}. We focus on both standard logistic regression, and  generalized additive models based on P-splines. Moreover, the random forest as a representative from the field of machine learning is shortly presented. Then, in Section~\ref{sec:IML} several interpretable machine learning methods are explained, including partial dependence plots (PDP) and individual conditional expectation (ICE) plots. Finally,  in Section~\ref{secperfmeas} various performance measures are defined.

\subsection{Statistical and machine learning methods}
\label{sec:reg}

In this section, we introduce the classic logistic regression model for binary outcomes, followed by its extension to non-linear effects via spline-based approaches.

\label{sec:regression}
\subsubsection*{Logistic regression}

Given observations $(y_i ,x_{i1},\ldots,x_{ip})$ for $i =1,\ldots,n$ tennis
matches, 
$$
\pi_i = P(y_i = 1 | x_{i1},\ldots,x_{ip}) = E[y_i | x_{i1},\ldots,x_{ip}]
$$
is the (conditional) probability for  $y_i = 1$, i.e.\ Player 1 winning the match, given covariate values $x_{i1},\ldots,x_{ip}$.

 We further specify a strictly monotonically increasing \textit{response function} 
${h\colon \mathds{R}\to [0,1]}$,
\begin{equation}
\pi_i = h(\eta_i) = h(\beta_0 + \beta_1 x_{i1} +\cdots+ \beta_p x_{ip})\,,\label{eq:logit}
    \end{equation}
which relates the linear predictor $\eta_i$ to $\pi_i$.

The logistic regression model, which uses the sigmoid function as response function, is the most famous candidate within the framework of Generalized Linear Models (GLMs).

The corresponding estimates $\hat\beta_0,\ldots,\hat\beta_p$ are obtained by numerical maximization of the underlying log-likelihood, e.g.\ by using Fisher scoring or the Newton-Raphson algorithms (see, e.g., \citealp{NelWed:72}). 
This approach is implemented in the \texttt{glm} function from the \texttt{stats} package in R.
For more details on GLMs, see \citet{FahTut:2001}.

\subsubsection*{Spline-based approaches}

In the classic logistic regression model introduced above, the covariates effects are
strictly linear, see equation~\eqref{eq:logit}.
However, in practice also non-linear influences might be relevant. In order to model these appropriately and flexibly, the GLM from above be extended to a so-called Generalized Additive Model (GAM; \citealp{wood2017generalized}).
For this purpose, so-called \textit{splines} can be used. In this work, we focus on \textit{B-splines} (see, e.g., \citealp{eilers2021practical}).

So instead of linear effects like the $\beta_j x_{ij}$ in equation~\eqref{eq:logit}, with B-splines a non-linear effect $f(x)$  of a metric predictor can be represented as 
$$
f(x) = \sum_{j = 1}^d \gamma_j B_j(x)\,,
$$
where $B_j(x)$ are different B-spline basis functions, $d$ denotes the number of basis functions used, and $\gamma_j$ the corresponding spline coefficients.
As an unpenalized estimation of a non-linear B-spline effect often overfits, 
typically the non-linear effect is smoothed by using {\it penalized B-splines}, i.e. {\it P-splines}.

Beside the problem of potential overfitting, the goodness-of-fit of the B-spline approach depends on the number of selected nodes. To avoid this problem, various penalization methods exist in the form of P-splines. Here, a penalized estimation criterion, which is extended by a penalty term, is used instead of the usual estimation criterion. For P-splines based on B-splines see, e.g., \citet{EilMar:96}. 
For this approach, we rely on the \texttt{gam} function from the \texttt{mgcv} package \citep{Wood:2017} in R. Note that such P-spline based approaches have also been used in \citet{buhamra:2024} and \citet{buhamra:2025}.

\subsubsection*{Random forest}

In the following, a random forest will be used for comparison with the linear and non-linear regression approaches introduced above.
This method is based on a (typically large) ensemble of trees, which were introduced by \citet{breiman1984classification}. However, as     
individual trees suffer from instability \citep{breiman1996heuristics}
an ensemble method called {\it bootstrapping and aggregating} (bagging; \citealp{breiman1996bagging}) was introduced, which in general improves the predictive performance compared to a single regression tree and is rather easy to implement.

A {\it random forest} aggregates multiple decision trees to enhance prediction accuracy \citep{breiman2001random}. The key idea is that combining uncorrelated models (individual trees) reduces variance and improves predictions compared to using a single model. In this manuscript,  classifications trees are considered in  the {\it random forest} (where the most frequent class among the trees determines the outcome), as our target variable is binary. For this purpose, the \texttt{ranger} package in R by \cite{ranger:2017} is used for fitting the random forest models. Also, the two hyperparameters \texttt{mtry} and\texttt{ntree} are quite important. For optimizing \texttt{mtry}, 10-fold cross-validation is used, and \texttt{ntree} is set to $400$. Further details about these parameters can be found in \citet{buhamra:2025}

\subsection{Interpretable machine learning} \label{sec:IML}

In the following, we discuss interpretable machine learning (IML) methods such as \textit{partial dependence plots} (PDP; see \citealp{friedman2001greedy}) and \textit{individual conditional expectation} (ICE) plots \citep{goldstein2015peeking} in more detail. These methods aim to enhance the interpretability of complex, black box-type machine learning models. Particularly, they can be applied to such black box models to provide explanations for individual predictions or overall model behavior. For this purpose, the implementations from the R packages \texttt{pdp} by \cite{greenwell2017pdp} and \texttt{ggplot2} by \cite{wickham2016package} are used (the latter one being generally applied for plotting). A nice overview of standard approaches for IML can be found in \citet{molnar2020interpretable}.

\subsubsection*{Partial Dependence Plot (PDP)}

Following \cite{molnar2020interpretable}, the partial dependence plot (or PDP)  illustrates the marginal effect of one or two features on the predicted outcome of a machine learning model \citep{friedman2001greedy}. It can reveal whether the relationship between a feature and the target is linear, monotonic, or more complex. The partial dependence function for regression is defined as follows
\begin{equation*}
\hat{f}_S(\mathbf{x}_S) = \mathds{E}_{\mathbf{x}_C}[\hat{f}(\mathbf{x}_S, \mathbf{x}_C)] = \int \hat{f}(\mathbf{x}_S, \mathbf{x}_C) \, dP(\mathbf{x}_C)\,,
\end{equation*}
where $P(\cdot)$ is the distribution of the features in set \( C \), \( \mathbf{x}_S \) are the features for which the partial dependence function is plotted, while \( \mathbf{x}_C \) represents the other features used in the machine learning model \( \hat{f}(\cdot) \), which are considered as random variables in this context. Typically, the set \( S \) contains only one or two features, namely those whose effect on the prediction one aims to understand. The feature vectors \( \mathbf{x}_S \) and \( \mathbf{x}_C \) together form the complete feature space \( \mathbf{x} \). Partial dependence operates by marginalizing the model's output over the distribution of the features $P(\cdot)$ in set \( C \), allowing the function to reveal the relationship between the features in \( S \) and the predicted outcome. By doing so, we obtain a function depending solely on the features in \( S \), while still accounting for interactions with other features. The partial function \( \hat{f}_S \) is given by 
\begin{equation*}
\hat{f}_{S}(\mathbf{x}_S) = \frac{1}{n} \sum_{i=1}^{n} \hat{f}( \mathbf{x}_{S},\mathbf{x}_{i,C})\,,
\end{equation*}
i.e.\ it is estimated by calculating averages on the training data. The partial function shows the average marginal effect on the prediction for given values of the features in~$S$. Here, $\mathbf{x}_{i,C}$ denotes the actual values from the dataset for the features not of interest, and 
$n$ represents the number of instances in the dataset. A key assumption of the PDP is that the features in \( C \) are uncorrelated with those in 
\( S \). If this assumption does not hold, the calculated averages may include data points that are highly unlikely or even impossible. Further details can be found in \cite{molnar2020interpretable}.

\subsubsection*{Individual Conditional Expectation (ICE)}

The counterpart to a PDP \citep{friedman2001greedy}, which illustrates the average effect of a feature, is referred to as individual conditional expectation (ICE) plot and is applied for individual data instances. The approach was first introduced by \cite{goldstein2015peeking}. ICE plots are used in machine learning to analyze the relationship between a feature and the predicted outcome for individual instances within a dataset. Unlike PDPs, which focus on the average effect of a feature, ICE plots offer insight into how changes in a specific feature or feature set impact the model's predictions for individual instances. Each line in an ICE plot represents the predicted outcome for a single instance as the feature value varies, revealing the variability and heterogeneity in feature effects across different instances. This helps identifying interactions between features,  understanding complex model behaviors, and detecting outliers, thereby improving model interpretability and transparency.

\subsection{Performance measures}
\label{secperfmeas}

In the following, performance measures are defined which we use to select the best model based on the predictive performance with regard to those measures (see also \citealp{buhamra:2024}, and \citealp {buhamra:2025}). Let $\tilde y_1,\ldots,\tilde y_n$ denote the true binary outcomes of a set of $n$ matches, i.e., $\tilde y_i \in \{0,1\}, i = 1,\ldots,n$. Moreover, let $\hat{\pi}_{i1} =: \hat{\pi}_i$ denote the probability, predicted by a certain model, that player~1 wins match~$i$. Then, the probability that player~2 wins the match is directly given by $\hat{\pi}_{i2} = 1 - \hat{\pi}_{i1}$. 

\subsubsection*{Classification rate}

The (mean) \textit{classification rate}
is given by the proportion of matches
correctly predicted by a certain model, i.e.\ 
$$
\frac{1}{n} \sum_{i=1}^n \mathbb{1} (\tilde y_i = \hat y_i)\mbox{, where\quad} \hat y_i = \begin{cases}
	1,&\text{$\hat{\pi}_i > 0.5$}\\
	0,&\text{$\hat{\pi}_i\le 0.5$}
\end{cases}\,,
$$
see, e.g., \citet{SchauGroll2018}. Hence, large values indicate a good predictive performance.

\subsubsection*{Predictive Bernoulli likelihood}

Following again \citet{SchauGroll2018},
the (mean) \textit{predictive Bernoulli likelihood} is based on the predicted probability $\hat \pi_i$ for
the true outcome $\tilde y_i$. For $n$ observations it is defined as

$$
\frac{1}{n} \sum_{i=1}^n{\hat \pi_i}^{\tilde y_i} ({1-\hat \pi_i})^{1-\tilde y_i}\,,
$$
Once again, large values indicate  good predictive performance.

\subsubsection*{Brier score}

The \textit{Brier score} \citep{Brier:50} is based on the squared distances between the 
predicted probability $\hat \pi_i$ and the actual (binary) output $\tilde y_i$ from match $i$, and is defined as
$$
\frac{1}{n} \sum_{i=1}^n ({\hat \pi_i - \tilde y_i})^2\,.
$$
It is an error measure and, hence, low values indicate a good predictive performance.

\section{Results}\label{sec:five}

In the next section, the predictive power of the enhanced variables is presented for both regression and machine learning approaches. For this purpose, we use an expanding window (EW), a rolling window (RW) as well as  a leave-one-tournament-out cross-validation (CV) approach. The best models are identified  with respect to the previously defined performance measures. 
Additionally, we provide better interpretability of the machine learning approach, i.e.\ the random forest model, by using IML tools such as partial dependence plot (PDP) and individual conditional expectation (ICE). All calculations and evaluations were performed using the statistical programming software R \citep{RDev:2024}.

\subsection{Enhanced variables predictive power}\label{sec:EnhanceVar}

To investigate the predictive potential of so-called `statistically enhanced covariates' in more detail, certain promising variables are considered, such as \textit{Elo, Age.30} and \textit{Age.int}, along with two conventional covariates (\textit{Rank} or \textit{Points}). All possible combinations of these covariates result in a total of 21 models for each of our proposed approaches, including linear effects, non-linear effects (splines), and the random forest. The results are given in Tables~\ref{t1} and \ref{t2}, with the best performers highlighted in bold.  For the expanding window approach, results are presented here in detail, as this method clearly identifies the top-performing models among the 21 proposed. The results for the leave-one-tournament-out cross-validation and rolling window approaches are included in the appendix.

\subsubsection*{Expanding window validation}

We validated all proposed 21 models with respect to their predictive performance on new, unseen test data. The validation is performed using an expanding window forecasting approach, i.e., each time one of the remaining tournaments is used as the test data set in chronological order, and the training data set is constantly updated and enlarged. This scheme has already been previously used in \cite{buhamra:2025} and can be explained as follows:

\begin{enumerate}
\item  First, all tournaments prior to 2022 are used as the training data set. Then, all models are fitted. Based on those, predictions are
derived for the 2022 Australian Open matches, as this was the 1$^{st}$ Grand Slam tournament in 2022.
\item  Then, the training data is updated, by adding the matches of the Australian Open 2022. Then, the
models are fitted again o the extended training data, and predictions are made for the French Open 2022, which is the 2$^{nd}$ Grand
Slam tournament in 2022.
\item Next, the matches of the French Open 2022 are added to the training data set and the models are fitted
again. Based on those fits, Wimbledon 2022 is predicted.
\item Then, the Wimbledon 2022 matches will be added to the training data set, and again, the models are
fitted and predictions are made for the final Grand Slam tournament in 2022, the US Open 2022.
\end{enumerate}
Finally, the prediction results for all four tournaments are compared with the actual match outcomes, and the corresponding performance measures are calculated .
Table~\ref{t1} presents the predictive performance for the regression models with linear and non-linear effects, resulting from all 21 possible combinations of the statistically enhanced variables.
Moreover, in Table~\ref{t2}, the same covariate combinations are presented, but with respect to the random forest method.

\textbf{Linear model:} The classification rates range from 0.737 to 0.795. The best performance was achieved by the linear regression model including \textit{Points}, \textit{Rank}, and \textit{Elo}, with a value of 0.795. For the predictive likelihood, the best value is 0.701, achieved by the model including the covariates \textit{Points}, \textit{Elo} and \textit{Age.int}.
The lowest Brier score, hence the best, is 0.153 and is delivered by eight different models.

\textbf{Splines model:} Unlike the models with linear effects, for the non-linear models are clear winning model can be identified based on the enhanced variables \textit{Elo} and \textit{Age.30}. Specifically, the splines model including covariates \textit{Elo} and \textit{Age.30} demonstrates the best performance with respect to all proposed performance measures. This model achieved a classification rate of 0.792, and for the predictive likelihood and  Brier score, the corresponding values are 0.703 and 0.149, respectively. Additionally, it should be noted that with respect to the classification rate, the model incorporating \textit{Points}, \textit{Elo} and \textit{Age.30} also yielded identical results to the winning model, namely the value 0.792.

\textbf{Random forest:} Similar to the case of the spline-based approaches, the winning model among the random forests can be clearly identified. The results show that the random forest based on \textit{Points}, \textit{Rank}, \textit{Age.30}, and \textit{Elo} covariates outperforms all other models.  It yielded a classification rate of 0.820, a predictive likelihood of 0.667, and a Brier score of 0.151.

Overall, if we considered both the proposed model types (i.e., linear regression, spline-based regression and random forests) and different forecasting strategies (such as EW, CV, and RW), certain models consistently prevail when these enhanced variables are present.
The results of the leave-one-tournament-out CV approach for the regression approaches (with linear and non-linear effects), as well as for the random forest are presented in Tables~\ref{t4} and ~\ref{t5}, respectively, in the appendix. Also, the corresponding results of the rolling window approach are available in appendix (see Tables~\ref{t6} and ~\ref{t7}). In general, across all the tables presented in the appendix, the results suggest a nearly identical trend: prediction accuracy improves when at least one of these enhanced variables (i.e., \textit{Elo}, \textit{Age.30}, \textit{Age.int}) is included.

\begin{table}[H]
\fontsize{10}{13}\selectfont
\centering

	\caption{ Results of the expanding window approach incorporating the statistically enhanced covariates for the regression model class with linear and non-linear effects; best results are highlighted in bold font.}

	\label{t1}
	\begin{tabular}{llccc@{\quad }c@{\;(}c@{)} }
	\multirow{2}{*}{}& Specific & Class. & Likeli- & \multicolumn{1}{c}{Brier} \\
		& models & rate & hood & \multicolumn{1}{c}{score} \\ [\smallskipamount]
		\hline
		\multirow{21}{*}{Linear}& Points &  0.754   & 0.639   & 0.178      \\
            & Elo & 0.782   &  0.695    &    $\mathbf{0.153}$   \\
            & Rank &  0.754  &   0.642   &   0.177      \\
             & Points, Rank & 0.754 & 0.661 & 0.171  \\
		
              & Points, Elo & 0.782  &  0.695   &  $\mathbf{0.153}$       \\
            & Rank, Elo &  0.775   & 0.696     & $\mathbf{0.153}$        \\
             & Elo, Age.30  & 0.778    & 0.697     &  $\mathbf{0.153} $    \\
             &Rank, Age.30 &  0.747      & 0.649       &   0.174     \\
             & Points, Age.30  & 0.758     &  0.643     & 0.176       \\
              & Elo, Age.int  &   0.775  &  0.696    &   0.154            \\
             &Rank, Age.int & 0.737  &   0.647     &  0.175      \\
             & Points, Age.int  &  0.737    & 0.642      & 0.177     \\
             
		& Points, Rank, Elo & $\mathbf{0.795}$ & 0.696 & $\mathbf{0.153}$  \\
		& Points, Rank, Age.30 & $0.758$ & $0.666$ & $0.168$ \\
		& Points, Rank, Age.int & $0.754$ & $0.665$ & $0.169$ \\
		
            & Points, Elo, Age.30 & 0.785   &   0.696  &    $\mathbf{0.153}$               \\
            &  Elo, Rank, Age.30 & 0.782   &  0.698  & $\mathbf{0.153}$                  \\
            & Points, Elo, Age.int & 0.764  &  $\mathbf{0.701}$   &   0.1570                \\
            &  Elo, Rank, Age.int &  0.761    & 0.696     &  0.162               \\
		& Points, Rank, Age.30, Elo & $0.785$ & $0.698$ & $\mathbf{0.153}$ \\
		& Points, Rank, Age.int, Elo & $0.788$ & $0.697$ & $0.154$ \\
		
		\hline
		\multirow{21}{*}{Splines}&  Points &0.741     & 0.651  & 0.175     \\
            & Elo & 0.775   &  0.697    & 0.153      \\
            & Rank &  0.747  & 0.643     & 0.179        \\
             &Points, Rank & $0.744$ & $0.655$ & $0.174$  \\
		
            & Points, Elo & 0.775  & 0.698    &  0.153       \\
            & Rank, Elo &   0.778  &  0.696    &  0.156       \\
             & Elo, Age.30  & $\mathbf{0.792}$    & $ \mathbf{0.703} $   & $\mathbf{0.149}$              \\
             &Rank, Age.30 &   0.751     & 0.654       & 0.175       \\
             & Points, Age.30  & 0.761     & 0.659      & 0.171       \\
              & Elo, Age.int  &  0.768   & 0.699     & 0.157              \\
             &Rank, Age.int &    0.741    & 0.649       & 0.180       \\
             & Points, Age.int  & 0.734     & 0.655      & 0.178       \\
  
		& Points, Rank, Elo & $0.778$ & $0.697$ & $0.158$  \\
		& Points, Rank, Age.30 & $0.768$ & $0.665$ & $0.169$ \\
		& Points, Rank, Age.int & $0.741$ & $0.661$ & $0.175$ \\
		
             & Points, Elo, Age.30 & $\mathbf{0.792}$   &  0.701   &  $ 0.150$               \\
            &  Elo, Rank, Age.30 & 0.785     & 0.698     & 0.156                  \\
            & Points, Elo, Age.int &  0.764  &  0.701   & 0.157                  \\
            &  Elo, Rank, Age.int & 0.761     & 0.696     &  0.162               \\
		& Points, Rank, Age.30, Elo & $0.778$ & $0.697$ & $0.156$ \\
		& Points, Rank, Age.int, Elo & $0.761$ & $0.697$ & $0.163$ \\
\hline
\end{tabular}
\end{table}
\par \noindent

\begin{table}[H]
\fontsize{10}{13}\selectfont
\centering

	\caption{Results of the expanding window approach for random forest approach; best results are highlighted in bold font.}
	\label{t2}
	\begin{tabular}{llccc@{\quad }c@{\;(}c@{)} }
	\multirow{2}{*}{}& Specific & Class. & Likeli- & \multicolumn{1}{c}{Brier} \\
		& models & rate & hood & \multicolumn{1}{c}{score} \\ [\smallskipamount]
  
		\hline
		\multirow{21}{*}{Random forest} &  Points &  0.773   & 0.596   & 0.189      \\
            & Elo & 0.800   & 0.615     & 0.174      \\
            & Rank &  0.784  &   0.596   &  0.187       \\
            &Points, Rank & $0.779$ & $0.616$ & $0.179$ \\
           
            & Points, Elo & 0.799  & 0.632    & 0.168        \\
            & Rank, Elo &   0.804  &   0.634   & 0.164        \\
             & Elo, Age.30  & 0.793    & 0.616     &  0.173             \\
             &Rank, Age.30 &    0.797    &  0.596      & 0.186       \\
             & Points, Age.30  & 0.778     & 0.598      &  0.187      \\
              & Elo, Age.int  &  0.804   &  0.618    & 0.172              \\
             &Rank, Age.int &   0.793     & 0.595       & 0.187       \\
             & Points, Age.int  &   0.778  & 0.595      & 0.188       \\
		& Points, Rank, Elo & $0.807$ & $0.638$ & $0.164$  \\
		& Points, Rank, Age.30 & $0.791$ & $0.616$ & $0.178$ \\
		& Points, Rank, Age.int & $0.788$ & $0.616$ & $0.178$ \\
		
            & Points, Elo, Age.30 & 0.779   & 0.631    &     0.172              \\
            &  Elo, Rank, Age.30 &  0.796    & 0.632     & 0.166                 \\
            & Points, Elo, Age.int &  0.808  & 0.632   &      0.162             \\
            &  Elo, Rank, Age.int &  0.794    &  0.634    &       0.165          \\
		& Points, Rank, Age.30, Elo & $\mathbf{0.820}$ & $\mathbf{0.667}$ & $\mathbf{0.151}$ \\
		& Points, Rank, Age.int, Elo & $0.800$ & $0.637$ & $0.166$ \\
		\hline
  
	\end{tabular}
\end{table}
\par \noindent

\subsection{Model interpretation}
\label{MI}

In this section, our interpretation is developed based on the refitting of the top-performing linear, spline, and random forest models introduced in Section~\ref{sec:EnhanceVar}. Consequently, we provide deeper insights for these models. Additionally, interpretable machine learning tools such as partial dependence plot (PDP) and individual conditional expectation (ICE) are used to enhance our understanding of the random forest model.

\subsubsection{Linear model}

The linear model can be directly interpreted based on the $p$ estimated regression coefficients $\hat\beta_j$ from Equation~\ref{eq:logit}.
Their values indicate how much the outcome variable is expected to change when the corresponding predictor variable changes by one unit, assuming all other covariates in the model are held constant.
Hence, they allow direct interpretation of both strength and direction of the relationship between the predictors and the outcome variable.
Table~\ref{t33} shows the coefficient estimates for the best candidate linear model based on the results from the previous Section~\ref{sec:EnhanceVar}. It incorporates \textit{Points}, \textit{Rank} as conventional covariates, and \textit{Elo} as  enhanced covariate.

\begin{table}[H]
\fontsize{10}{13}\selectfont
\centering
\caption{Estimated coefficients for the candidate linear model based on the results from Table \ref{t1}}
\label{t33}
\begin{tabular}{@{\hspace{8pt}}p{0.02cm}@{\hspace{8pt}} @{\hspace{8pt}}p{1.5cm}@{\hspace{8pt}} @{\hspace{8pt}}r@{\hspace{8pt}}}
    & Variables & $\hat\beta_j$\phantom{mm} \\ [\smallskipamount]
    \hline
    & Rank & -0.0012  \phantom{mm}  \\    
    & Elo & 0.0042 \phantom{mm}  \\ 
    & Points & 0.0001   \phantom{mm} \\ 
    \hline
\end{tabular}
\end{table}

For the \textit{Rank} difference variable, the negative sign indicates that as this variable increases (resulting in a larger numerical rank, which corresponds to a lower-ranked first-named player), the linear predictor decreases by approximately 0.0012 units. Hence, due to the the used (logistic) response function, also the winning probability for the respective player decreases.

Analogously, for every one-unit increase in the \textit{Elo} rating difference, the linear predictor exhibits an increase of approximately 0.0042 units. Hence, also  the winning probability for the respective first-named player increase. In practical terms, this means that  larger \textit{Elo} ratings are associated with larger probabilities of winning a tennis match, as the positive sign of the coefficient estimate indicates a positive relationship.

The same positive relationship also holds for the predictor variable \textit{Points}. For every one-unit increase in the difference in points between players (again, from the perspective of the first-named player), the model predicts an increase of approximately 0.0001 units in the linear predictor and, hence, also an increase of the respective player's winning probability.

\subsubsection{Splines}

Graphical illustrations of spline effects are a useful way to understand and visualize potential non-linear effects of predictors on the response variable. The spline-based approaches from Section~\ref{sec:reg} are able to capture non-linear relationships and allow for a flexible, semi-parametric form of regression using smooth functions (splines) for predictors. Corresponding ploting functions provide interpretable visualizations. In this context, the \texttt{gam} function from the \texttt{mgcv} package \citep{wood2017generalized} in R is used.

Figure~\ref{fig:GAM_plot} displays the effects for the covariates \textit{Elo}  and \textit{Age.30}, which actually appear to be linear.
The graphs show positive effects for both features, i.e.\ larger \textit{Elo} ratings associated with larger winning probabilities.
The \textit{Age.30} difference has a slightly positive effect, which however is not significant taking the point-wise confidence band into account.

\begin{figure}[h]
    \centering
    \includegraphics[width=0.8\textwidth]{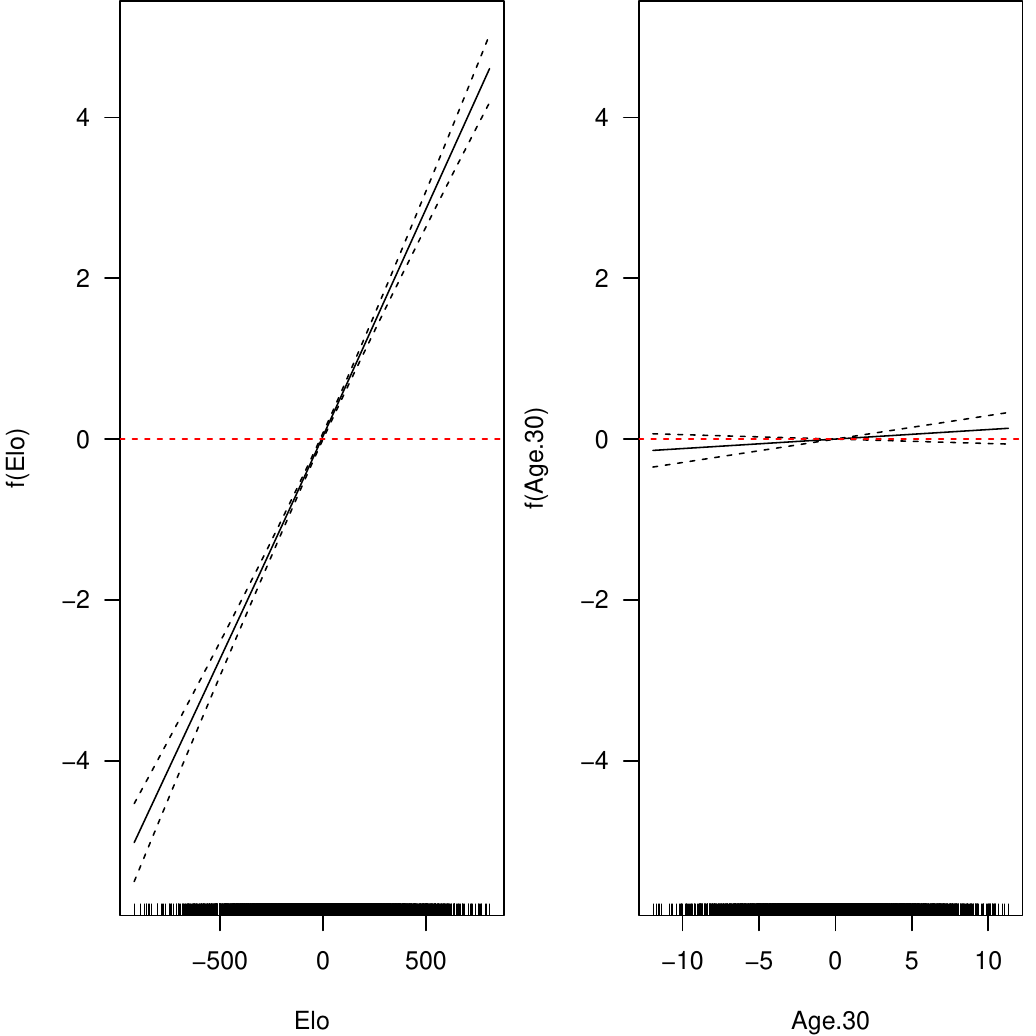}
    \caption{Spline effects for the enhanced covariates \textit{Elo} and \textit{Age.30}}
    \label{fig:GAM_plot}
\end{figure}

\subsubsection {Random forest with interpretable machine learning}

Next, the results for the random forest model with the best predictive performance from Section~\ref{sec:EnhanceVar} are discussed.
This model, which included \textit{Points}, \textit{Rank}, \textit{Age.30}, and \textit{Elo} as covariates, is refitted to our entire data set. Then, both partial dependence and Individual Conditional Expectation (ICE) plots are presented for better interpretability and visualization of the respective complex model.

\begin{figure}[h]
    \centering
    \includegraphics[width=0.8\textwidth]{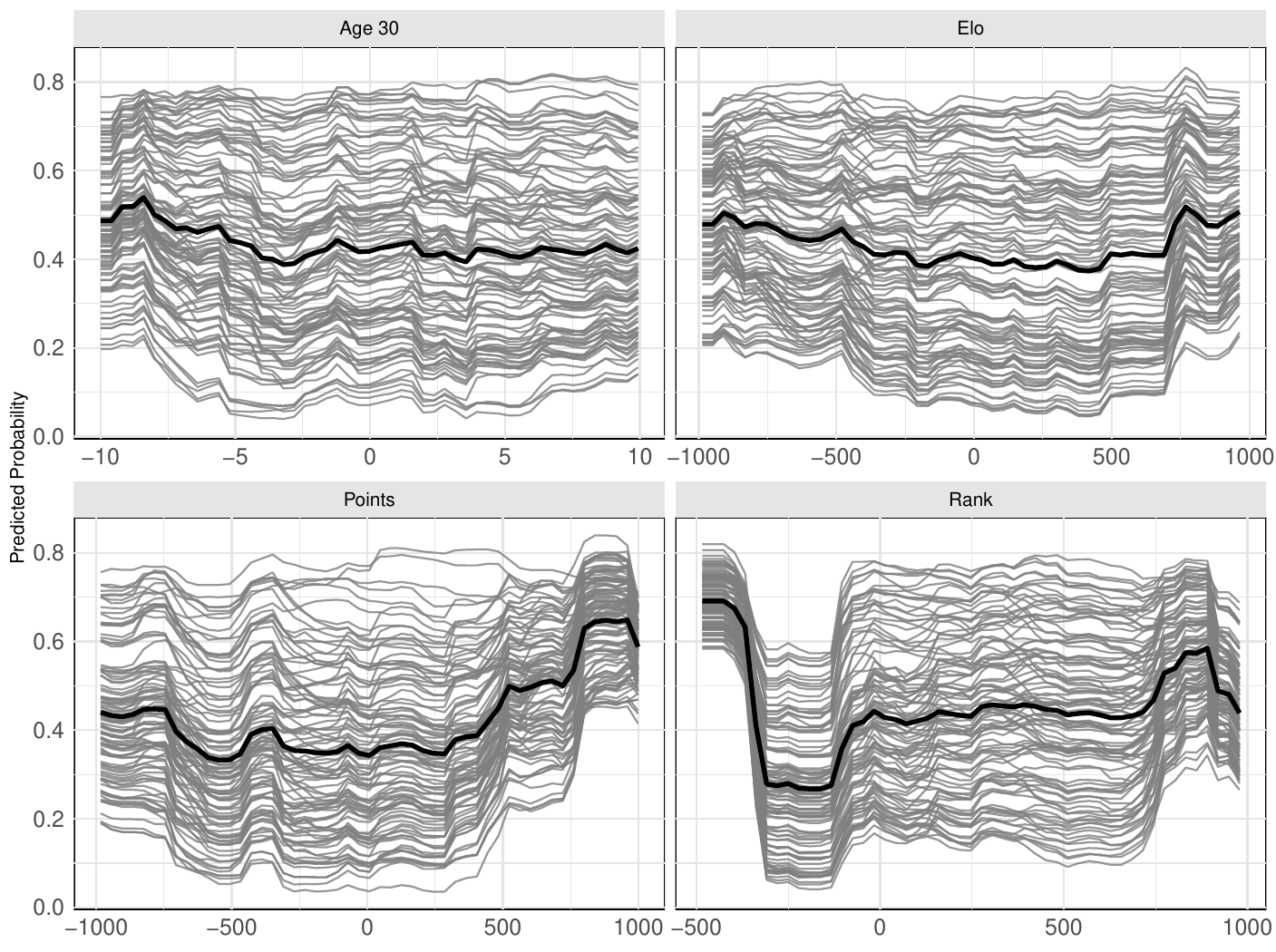}
    \caption{Combined PDP (thick black line) and ICE plots (grey lines) for the random forest model including covariates \textit{Age.30}, \textit{Elo}, \textit{Points} and \textit{Rank}}
    \label{fig:Combine_ICE_PDP}
\end{figure}

\begin{figure}[h!]
    \centering
    \includegraphics[width=0.7\textwidth]{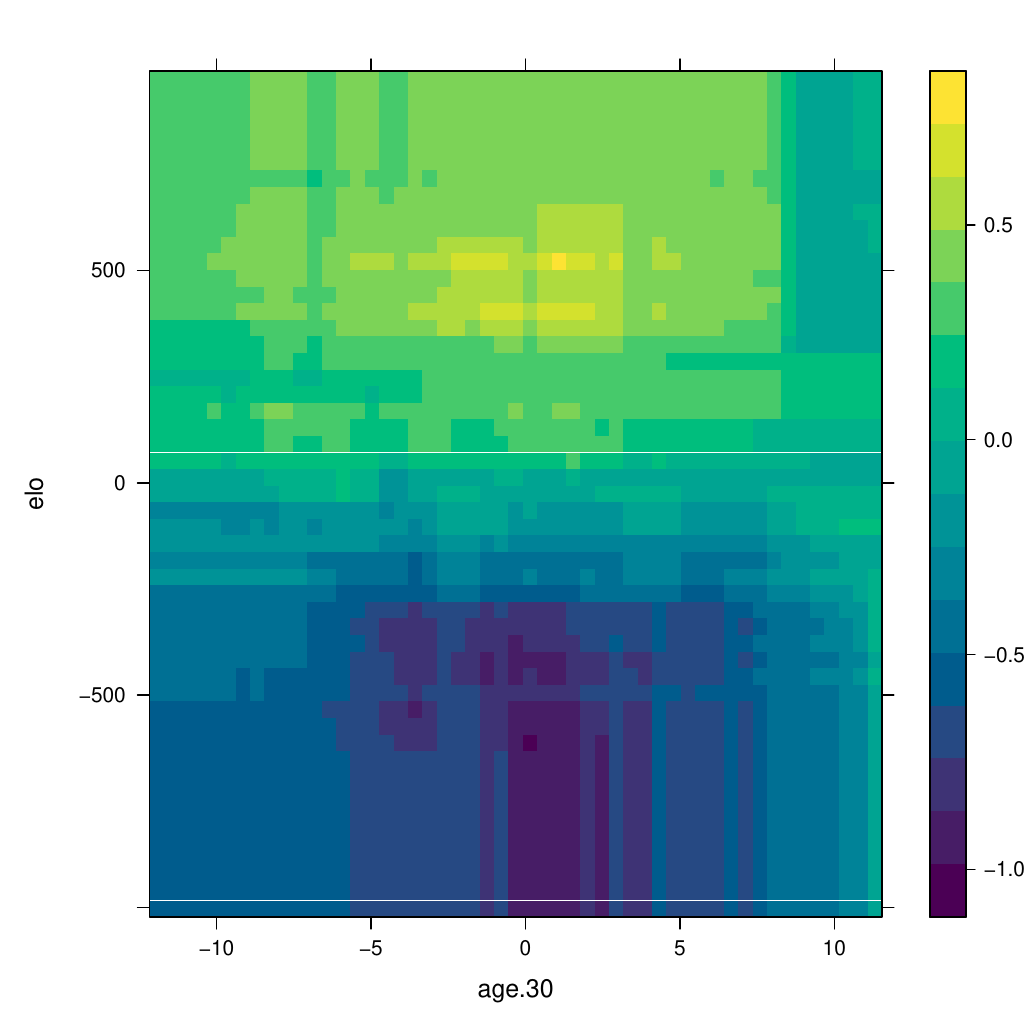}
    \caption{Heat map for the random forest model including \textit{Elo} and \textit{Age.30} as covariates}
    \label{fig:heatmap_elo_Age}
\end{figure}

Since ICE plots—the individual equivalent of Partial Dependence Plots (PDPs)—are considered crucial in our case, the combined ICE and PDP plots were used to provide a comprehensive visualization of the relationship between predictor variables and the predicted outcome.

In Figure~\ref{fig:Combine_ICE_PDP} we generated a combined ICE and PDP. Each ICE line represents how the predicted winning probability changes with the predictor value for a single observation. The spread of the lines indicates variability in the effect of the predictor across different observations. The thick black line represents the PDP, showing the average effect of the respective predictor on the predicted outcome. This PDP line provides context to the ICE lines by highlighting the overall trend across all observations. 

For instance, in the \textit{Rank} panel, we can observe how the predicted outcome is first very strongly decreasing and then slowly increasing. Moreover, in the \textit{Rank} panel, the spread of grey lines illustrates variability in how an increasing \textit{Rank} difference affects different observations. After a steep decrease, the general trend of the black line indicates a almost constant, very slightly increasing relationship between \textit{Rank} and the predicted probability of winning. This means that as the \textit{Rank} difference increases (i.e., the first-named player is ranked higher, meaning worse performance, when the rank of the second player is kept constant), the winning probability first strongly decreases, then increases a bit and finally remains almost constant, before it exhibits a small hill at the end.

In the \textit{Elo} panel, the PDP line shows a slightly negative trend, suggesting that larger \textit{Elo} differences first decrease the predicted probability of winning, before increasing again. 
For the \textit{Points}, upward slope in the PDP  suggests that accumulating more points correlates positively with increased chances of winning matches. Variability among ICE lines indicate that while most players benefit from accumulating points, some may experience diminishing returns at higher point levels or face challenges against specific opponents. 
The \textit{Age.30} panel shows significant variability in individual responses, as indicated by the spread of the ICE lines. The ICE lines highlight differences among individual responses: some (first-named) players with an age closer to the optimal one of 30 years outperform the (second-named) player substantially. In addition, we can see a slightly decreasing trend of the \textit{Age.30} difference, which then seems to then remain rather constant for small or positive differences. 

Overall, this combined plot effectively conveys both the average influence of each predictor (PDP) and its varying impact on individual observations (ICE). It allows for precise interpretation of how predictors influence the outcome across different levels and individuals.

Finally, in Figure~\ref{fig:heatmap_elo_Age}, a heatmap plot for the partial dependency of \textit{Elo} and \textit{Age.30} is presented. The plot examines how the linear predictor (and, hence, the wining probability) changes jointly with \textit{Age.30} and \textit{Elo}. If the color changes are not uniform across the heat map, this suggests a complex interaction between \textit{Elo} and \textit{Age.30}. For example, larger \textit{Elo} have a stronger impact on the predictor for certain age ranges than for others (e.g., players with an age closer to 30  benefit more from a larger Elo ratings than younger or older players).

For example, a region that transitions from blue to yellow as \textit{Elo} increases indicates that higher \textit{Elo} scores are associated with higher winning probabilities, especially when \textit{Age.30} is within a certain range.

\section{Summary and overview} \label{sec:six}

In this work, we compared different model approaches for modeling tennis matches in Grand Slam tournaments focusing on two main aspects: statistically enhanced covariates and interpretable machine learning tools. First, we demonstrated how these enhanced covariates can be applied in the context of sports analytics to improve predictive modeling performance and gain insights from sports data. Then, to better understand the interpretation of complex ML models, exemplarily for a random forest, we presented partial dependence plots (PDPs) to visualize the average partial relationship between the predicted response and one or more features, along with individual conditional expectation (ICE) plots, a tool for visualizing the model estimated by any supervised learning algorithm. For this purpose, we used a data set provided and analyzed in \citet{buhamra:2024, buhamra:2025}, which was compiled based on the R package \texttt{deuce} \citep{Kovalchik2018deuce}. It contains information on 5,013 matches in 47 men's Grand Slam tournaments from the years 2011-2022. It also includes covariate information on the age difference of both players (\textit{Age}), the difference in their ranking positions (\textit{Rank}) and ranking points (\textit{Points}), in Elo numbers (\textit{Elo}), as well as the two additional age-based variables, \textit{Age.30} and \textit{Age.int}, which were constructed such that they take into account that the optimal age of a tennis player is between 28 and 32 years.

Different regression models, which were already considered in \citet{buhamra:2024, buhamra:2025}, were compared with machine learning approaches, in particular a random forest model, for modeling and predicting tennis matches. Since there are only two possible outcomes in tennis (win or loss), all models were based on a binary outcome and thus focused on modeling the probability of the first-named player winning.

The different modeling approaches were compared with respect to their prediction performance on unseen matches via an expanding window strategy.  The following regression and ML approaches were included in this comparison:
\begin{itemize}
	\item Logistic regression with linear effects: all possible combinations of the three enhanced covariates along with the conventional variables \textit{Point} and \textit{Rank} were considered. This resulted in 21 models.
	\item Logistic regression with non-linear effects (splines): Again, the same 21 combinations of enhanced and conventional covariates were considered.
     \item A random forest model as a machine learning approach: Also here, the same combinations of enhanced and conventional covariates  were considered, resulting in 21 models.
\end{itemize}
Via the expanding window approach, the models were compared in terms of  classification rate,  predictive Bernoulli likelihood and  Brier score. Since each approach resulted in 21 different models, the model with the best predictive performance measures was selected in each case. Overall, the values vary between the different approaches and over proposed performance measures. The random forest model, including the covariates \textit{Points}, \textit{Rank}, \textit{Age.30}, and \textit{Elo} achieved the best classification rate among all other models with a value of 0.8202. In contrast, the spline-based regression model based on \textit{Elo} and \textit{Age.30} only yielded the best predictive performance in terms of predictive likelihood and Brier score compared to all other model approaches. Generally, one could say that models consistently perform better when at least one of the enhanced variables is included.

Additionally, we investigated a CV-type approach and a rolling window approach. The rolling window approached was based on a (varying) training dataset always containing 12 tournaments that were used to predict the outcome of the next tournament. The results for these two approaches are provided in the appendix. Principally, results are varying among the three different training-test-subdivision approaches, but generally led to similar results as the expanding window strategy. For example, the winning random forest model of the rolling window approach is similar to the one of the expanding window strategy, as both include \textit{Points}, \textit{Age.30}, \textit{Elo}, and \textit{Rank} as the best model predictors.

To gain a comprehensive understanding and interpretation of each approach, we analyze the coefficients of the linear regression model. Additionally, we employ spline graphs to visually represent and interpret the relationship between predictor variables and the response variable within the context of Generalized Additive Models (GAMs). These graphs can capture non-linear relationships, which is particularly advantageous for complex datasets where linear models may be insufficient. Furthermore, we introduce interpretable machine learning (IML) tools such as partial dependence plots (PDP) and individual conditional expectation (ICE) plots. These tools help in comprehending and interpreting the predictions made by complex (black box-type) machine learning models, in the present case a random forest model.

By examining PDP plots, we obtained insights into how each predictor variable affects the model’s predictions. For instance, the PDP for \textit{Points} exhibits  a general upward trend, indicating that earning more \textit{Points} has a positive effect on the predicted outcome. Similarly, the PDP for \textit{Age.30} suggests that being closer to the optimal age of 30 years is only slightly associated with larger winning probabilities.

In addition, a heat map visualization illustrating the joint effect of the two covariates \textit{Elo} and \textit{Age.30} on the predicted outcome is presented. By examining the color gradient and regions in the heat map, we can infer how changes in these two covariates influence the outcome, highlighting any non-linear interactions and dependencies between those.

In future research, additional IML methods, such as Accumulated Local Effects (ALE)  plots \citep{apley2020visualizing} and Local Interpretable Model-agnostic Explanations (LIME; \citealp{ribeiro2016should}) can also be used. Furthermore, one could investigate more complex machine learning models, such as deep learning approaches \citep{bishop1995neural, lecun2015deep}. Additionally, similar to soccer \citep{GroEtAl:WM2019}, one could also focus on tournament outcomes. For example, the probability of a certain player winning the tournament could be determined. This approach takes advantage of the fact that the tournament bracket is fully drawn before the start, allowing us to predict the earliest round in which two players could meet. Unlike in soccer, where group stage outcomes influence subsequent matches, this setup simplifies predictions. However, using only the match-specific betting odds for the first round presents a challenge. Models that exclude odds as covariates might be preferable, despite potentially lower prediction performance due to the significant influence of odds. Alternatively, models could be developed that use pre-tournament odds for each player to win the entire tournament, rather than odds for individual matches. Finally, additional statistically enhanced covariates could be produced. For example, similar to the historic match abilities for soccer teams developed by \citet{ley2019ranking}, such abilities could also be developed for tennis players. We are currently working on such an approach and plan to include those ability parameters into our models in the future.

\bibliographystyle{apalike}	
\bibliography{literature}

\newpage
\appendix 

\section{Appendix} 

\begin{table}[h!]
\fontsize{10}{13}\selectfont
\centering

	\caption{Results of the leave-one-tournament-out CV approach for regression model class; best results in bold font.}
	\label{t4}
	\begin{tabular}{llccc@{\quad }c@{\;(}c@{)} }
	\multirow{2}{*}{}& Specific & Class. & Likeli- & \multicolumn{1}{c}{Brier} \\
		& models & rate & hood & \multicolumn{1}{c}{score} \\ [\smallskipamount]
		\hline
		\multirow{21}{*}{Linear}& Points & $0.732 $   & $0.623 $  &    $0.185$   \\
            & Elo & $0.747 $  & $0.656$     &  $0.172$     \\
            & Rank &  0.732  & 0.592     & 0.199        \\
              &Points, Rank & $0.732$ & $0.634$ & $0.181$  \\
              & Points, Elo &    $0.748$  & $0.658$    &  $0.171$         \\
              & Rank, Elo & $\mathbf{0.749}$  &  $0.657 $   & $0.172 $       \\
             & Elo, Age.30  & $0.747 $   &   $0.656$   &     $0.172$          \\
             &Rank, Age.30 &    $0.731 $   &  $0.592 $     & $0.199 $      \\
             & Points, Age.30  &  $0.729$    &  $0.623$     & $0.185$       \\
              & Elo, Age.int  &  $0.747$   &  $0.656 $   &   $0.172 $           \\
             &Rank, Age.int &  $0.731 $     &   $ 0.592$    &  $0.199$     \\
             & Points, Age.int  & $ 0.731  $  &   $0.623 $   &   $ 0.185$    \\
		
		& Points, Rank, Elo & $0.747$ & $0.659$ & $\mathbf{0.170}$  \\
		& Points, Rank, Age.30 & $0.732$ & $0.634$ & $0.181$ \\
		& Points, Rank, Age.int & $0.733$ & $0.634$ & $0.181$ \\
		
            & Points, Elo, Age.30 &  $0.747$  & $0.658$    &  $0.171 $    \\
            &  Elo, Rank, Age.30 &  $ 0.747$   &  $0.657 $   & $ 0.172  $               \\
            & Points, Elo, Age.int &  $0.745$  &  $0.657 $  &   $  0.171$              \\
            &  Elo, Rank, Age.int &  $0.746 $   &  $\mathbf{0.656} $   &  $0.172 $              \\
		& Points, Rank, Age.30, Elo & $0.747$ & $\mathbf{0.659}$ & $\mathbf{0.170}$ \\
		& Points, Rank, Age.int, Elo & ${0.748}$ & $\mathbf{0.659}$ & $\mathbf{0.170}$ \\
		
		\hline

		\multirow{21}{*}{Splines}&  Points & 0.729    &  0.637  &  0.181     \\
            & Elo & 0.746   &  0.656    &   0.172    \\
            & Rank & 0.729   & 0.611     &  0.193       \\
             & Points, Rank & $0.732$ & $0.641$ & $0.179$  \\
            & Points, Elo &  0.748 &  $\mathbf{0.657} $  &  $\mathbf{0.171}$       \\
            & Rank, Elo &  $\mathbf{0.748}$   &  0.656    &  0.172       \\
             & Elo, Age.30  &  0.746   &  0.656    &   0.172            \\
             &Rank, Age.30 &    0.729    & 0.611       &  0.193      \\
             & Points, Age.30  &  0.729   & 0.637      &  0.181      \\
              & Elo, Age.int  &  0.745   &  0.656    &   0.172            \\
             &Rank, Age.int &    0.728    &   0.611     &    0.194    \\
             & Points, Age.int  &   0.729   &   0.637    &   0.181     \\
            
		& Points, Rank, Elo & $0.747$ & $\mathbf{0.657}$ & $\mathbf{0.171}$  \\
		& Points, Rank, Age.30 & $0.731$ & $0.641$ & $0.179$ \\
		& Points, Rank, Age.int & $0.730$ & $0.641$ & $0.179$ \\
		
            & Points, Elo, Age.30 & 0.746   & $\mathbf{0.657}$    & $\mathbf{0.171}$        \\
            &  Elo, Rank, Age.30 &   0.746   & 0.656     &  0.172                 \\
            & Points, Elo, Age.int &  0.745  & $ \mathbf{0.657}$   &   $\mathbf{0.171}$                \\
            &  Elo, Rank, Age.int &  0.746    &  0.656    &  0.172              \\
		& Points, Rank, Age.30, Elo & $0.745$ & $\mathbf{0.657}$ & $\mathbf{0.171}$ \\
		& Points, Rank, Age.int, Elo & $0.744$ & $\mathbf{0.657}$ & $\mathbf{0.171}$ \\

\hline
\end{tabular}
\end{table}
\par \noindent

\begin{table}[h]
\fontsize{10}{13}\selectfont
\centering
  \caption{Results of the leave-one-tournament-out CV approach for random forest models; best results in bold font.}
	\label{t5}
	\begin{tabular}{llccc@{\quad }c@{\;(}c@{)} }
	\multirow{2}{*}{}& Specific & Class. & Likeli- & \multicolumn{1}{c}{Brier} \\
		& models & rate & hood & \multicolumn{1}{c}{score} \\ [\smallskipamount]
		\hline
		\multirow{21}{*}{Random forest}& Points & 0.757    & 0.611   & 0.186      \\
            & Elo & 0.769   & 0.623     & 0.179      \\
            & Rank &   0.757 &  0.604    & 0.190        \\
  
            &Points, Rank & $0.759$ & $0.629$ & $0.181$ \\
            & Points, Elo &0.772   & 0.641    & $\mathbf{0.174}$        \\
            & Rank, Elo &  $\mathbf{0.773}  $ & 0.638    &  0.175       \\
             & Elo, Age.30  & 0.771    & 0.624     &  0.179             \\
             &Rank, Age.30 &   0.751     &  0.615      & 0.190       \\
             & Points, Age.30  &    0.755  &   0.608    & 0.187       \\
              & Elo, Age.int  &   0.769  &  0.620    &  0.179             \\
             &Rank, Age.int &    0.753    &  0.606      &   0.191     \\
             & Points, Age.int  & 0.754     & 0.608      & 0.187       \\
		& Points, Rank, Elo & $0.772$ & $\mathbf{0.645}$ & $\mathbf{0.174}$  \\
		& Points, Rank, Age.30 & $0.758$ & $0.626$ & $0.182$ \\
		& Points, Rank, Age.int & $0.760$ & $0.626$ & $0.182$ \\
		
            & Points, Elo, Age.30 & 0.769   &  0.638   & 0.175                  \\
            &  Elo, Rank, Age.30 &    0.772  & 0.635     & 0.176                  \\
            & Points, Elo, Age.int &  0.770  & 0.638    &  0.176                 \\
            &  Elo, Rank, Age.int &   0.770   & 0.636     &  0.176               \\
		& Points, Rank, Age.30, Elo  & $0.770$ & $0.643$ & $\mathbf{0.174}$ \\
		& Points, Rank, Age.int, Elo & $0.770$ & $0.643$ & $0.175$ \\
		\hline
  
	\end{tabular}
\end{table}
\par \noindent

\begin{table}[h!]
\fontsize{10}{13}\selectfont
\centering

	\caption{Results of the rolling window approach for regression model class; best results in bold font.}
	\label{t6}
	\begin{tabular}{llccc@{\quad }c@{\;(}c@{)} }
	\multirow{2}{*}{}& Specific & Class. & Likeli- & \multicolumn{1}{c}{Brier} \\
		& models & rate & hood & \multicolumn{1}{c}{score} \\ [\smallskipamount]
		\hline
		\multirow{21}{*}{Linear}& Points & 0.645    &  0.498  & $ 0.309$ \\
            & Elo & 0.630   &  0.496    &   0.330    \\
            & Rank & 0.645   &  $\mathbf{0.501}$  &  $\mathbf{0.293} $      \\
            &Points, Rank & $0.645$ & $0.499$ & $0.317$  \\
		
             & Points, Elo & 0.633  &  0.496   & 0.332        \\
            & Rank, Elo &   0.633  & 0.497     & 0.331        \\
             & Elo, Age.30  &   0.627  & 0.495     &  0.331             \\
             &Rank, Age.30 &   0.641     & $\mathbf{0.501}$      &  0.294      \\
             & Points, Age.30  &   0.642  & 0.498      & 0.309       \\
              & Elo, Age.int  & 0.631    &  0.495    & 0.331              \\
             &Rank, Age.int &   0.644     & $\mathbf{0.501}$      &  0.294      \\
             & Points, Age.int  & 0.646     &  0.498     &   0.309     \\
		& Points, Rank, Elo & $0.636$ & $0.497$ & $0.332$  \\
		& Points, Rank, Age.30 & $0.644$ & $0.499$ & $0.317$ \\
		& Points, Rank, Age.int & $\mathbf{0.647}$ & $0.499$ & $0.317$ \\
		
            & Points, Elo, Age.30 & 0.628   & 0.495    & 0.332                  \\
            &  Elo, Rank, Age.30 & 0.632     &  0.496    &  0.331                 \\
            & Points, Elo, Age.int &  0.631  & 0.495    &    0.332               \\
            &  Elo, Rank, Age.int &   0.635   & 0.496     & 0.331                \\
		& Points, Rank, Age.30, Elo & $0.633$ & $0.496$ & $0.332$ \\
		& Points, Rank, Age.int, Elo & $0.634$ & $0.496$ & $0.333$ \\
		
		\hline
		\multirow{21}{*}{Splines}& Points & $\mathbf{0.648}$    & 0.500   & 0.319      \\
            & Elo & 0.634   & 0.496     & 0.331      \\
            & Rank & 0.643   & $\mathbf{0.501}$     & $\mathbf{0.306}$        \\ 
            &Points, Rank & $0.646$ & $0.499$ & $0.323$  \\
		
             & Points, Elo & 0.633  & 0.496    & 0.332        \\
            & Rank, Elo &   0.634  & 0.497     & 0.330        \\
             & Elo, Age.30  &  0.633   & 0.496     &  0.331             \\
             &Rank, Age.30 &    0.645    &  $\mathbf{0.501}$ & $\mathbf{0.306}$       \\
             & Points, Age.30  &   0.644   & 0.499      & 0.320       \\
              & Elo, Age.int  & 0.632    & 0.496     &  0.331             \\
             &Rank, Age.int &  0.644      &  $\mathbf{0.501}$   & $\mathbf{0.306}$       \\
             & Points, Age.int  &  0.646    & 0.499      & 0.319       \\
		& Points, Rank, Elo & $0.633$ & $0.497$ & $0.332$  \\
		& Points, Rank, Age.30 & $0.643$ & $0.499$ & $0.323$ \\
		& Points, Rank, Age.int & $0.644$ & $0.499$ & $0.323$ \\
		
            & Points, Elo, Age.30 &  0.630  & 0.496    &   0.332                \\
            &  Elo, Rank, Age.30 &   0.634   & 0.496     &    0.331               \\
            & Points, Elo, Age.int & 0.633   &  0.496   &  0.333                 \\
            &  Elo, Rank, Age.int &  0.632    &  0.496    &   0.331              \\
		& Points, Rank, Age.30, Elo & $0.632$ & $0.496$ & $0.332$ \\
		& Points, Rank, Age.int, Elo & $0.632$ & $0.497$ & $0.332$ \\
\hline
\end{tabular}
\end{table}
\par \noindent

  \begin{table}[h]
\fontsize{10}{13}\selectfont
\centering
	\caption{Results of the rolling window approach for random forest models; best results in bold font.}
	\label{t7}
	\begin{tabular}{llccc@{\quad }c@{\;(}c@{)} }
	\multirow{2}{*}{}& Specific & Class. & Likeli- & \multicolumn{1}{c}{Brier} \\
		& models & rate & hood & \multicolumn{1}{c}{score} \\ [\smallskipamount]
		\hline
		\multirow{21}{*}{Random forest} & Points & 0.753    & 0.600   & 0.191      \\
            & Elo &  0.771  & 0.611     &  0.183     \\
            & Rank & 0.755   &   0.601   &   0.194      \\ 
           &Points, Rank & $0.756$ & $0.620$ & $0.185$ \\
           
            & Points, Elo &  0.771 &   0.628  &  0.177       \\
            & Rank, Elo &  0.773  & 0.627     & 0.179        \\
             & Elo, Age.30  & 0.771    &  0.610    &     0.184          \\
             &Rank, Age.30 &    0.748    &   0.600     &  0.195      \\
             & Points, Age.30  &   0.754   &  0.596     &   0.192     \\
              & Elo, Age.int  & 0.771    &  0.609    &    0.184           \\
             &Rank, Age.int &    0.752    &  0.600      & 0.195       \\
             & Points, Age.int  &  0.755    &  0.599     &  0.192      \\
		& Points, Rank, Elo & $0.773$ & $0.636$ & $0.177$  \\
		& Points, Rank, Age.30 & $0.757$ & $0.618$ & $0.186$ \\
		& Points, Rank, Age.int & $0.755$ & $0.616$ & $0.186$ \\
		
            & Points, Elo, Age.30 & 0.768    &  0.626   &  0.179                 \\
            &  Elo, Rank, Age.30 &   0.774   & 0.626     & 0.179                  \\
            & Points, Elo, Age.int &  0.765  &  0.625   &   0.179                \\
            &  Elo, Rank, Age.int &  0.769    &   0.625   &  0.179               \\
		& Points, Rank, Age.30, Elo & $\mathbf{0.789}$ & $\mathbf{0.659}$ & $\mathbf{0.165}$ \\
		& Points, Rank, Age.int, Elo & $0.771$ & $0.634$ & $0.177$ \\
		\hline
  
	\end{tabular}
\end{table}
\par \noindent

\end{document}